\documentclass[runningheads]{llncs}
\usepackage{makecell}
\usepackage{placeins}
\usepackage{multirow}
\usepackage{float}
\usepackage{graphicx}
\usepackage{graphicx}
\usepackage{wrapfig}
\usepackage{xspace}
\usepackage{color}
\usepackage{xcolor}
\usepackage{graphicx}
\usepackage{caption}
\usepackage{subcaption}
\usepackage{pgfplots}
\usetikzlibrary{patterns}
\usepackage{enumerate}
\usepackage{enumitem}
\usepackage{graphics}
\usepackage{subcaption}
\usepackage[skins]{tcolorbox}
\usepackage{amsmath}
\usepackage{appendix}
\usepackage{url}
\usepackage{color}
\usepackage{breqn}
\usepackage{pifont}
\usepackage{algpseudocode,caption}
\usepackage{float}
\usepackage{threeparttable}
\usepackage{booktabs}
\newcommand{\etal}{{\em et al.}\xspace}
\usepackage{xcolor}
\newcommand{\BfPara}[1]{{\vspace{0.3em}\noindent\bf#1.}\xspace}
\usepackage{xcolor}
\usepackage{colortbl}
\usepackage{tabularx}
\colorlet{lightgrey}{lightgray}
\usepackage{tikz}
\usepackage{pgf}
\usepackage{float}
\usepackage{authblk}

\usepackage{graphicx}
\definecolor{lavender}{rgb}{0.9, 0.9, 0.98}
\usepackage{tikz}

\usepackage{booktabs}
\usepackage{amsmath}

\usepackage[colorlinks=true, allcolors=blue]{hyperref}
\usepackage{array}
\newcolumntype{L}[1]{>{\raggedright\let\newline\\\arraybackslash\hspace{0pt}}m{#1}}
\newcolumntype{C}[1]{>{\centering\let\newline\\\arraybackslash\hspace{0pt}}m{#1}}
\newcolumntype{R}[1]{>{\raggedleft\let\newline\\\arraybackslash\hspace{0pt}}m{#1}}

\usepackage{makecell}
\newcommand{\xl}[1]{\Xhline{#1\arrayrulewidth}}

\begin{document}

\title{Dissecting the Infrastructure  Used in Web-based Cryptojacking: A Measurement Perspective}

\author[1]{Ayodeji Adeniran \and Kieran Human \and David Mohaisen}
\institute{University of Central Florida, Orlando. USA}

\maketitle

\begin{abstract}
This paper conducts a comprehensive examination of the infrastructure supporting cryptojacking operations. The analysis elucidates the methodologies, frameworks, and technologies malicious entities employ to misuse computational resources for unauthorized cryptocurrency mining. The investigation focuses on identifying websites serving as platforms for cryptojacking activities. A dataset of 887 websites, previously identified as cryptojacking sites, was compiled and analyzed to categorize the attacks and malicious activities observed. The study further delves into the DNS IP addresses, registrars, and name servers associated with hosting these websites to understand their structure and components. Various malware and illicit activities linked to these sites were identified, indicating the presence of unauthorized cryptocurrency mining via compromised sites. The findings highlight the vulnerability of website infrastructures to cryptojacking.

\end{abstract}

\section{Introduction}\label{sec:introduction} 

A cyberattack known as ``cryptojacking'' occurs when unauthorized individuals or entities utilize another's computer resources to mine cryptocurrencies. This attack type is also called cryptocurrency mining malware or malicious crypto mining. To maintain the security and validity of blockchain transactions, mining cryptocurrencies requires solving intricate mathematical problems, which demand a considerable amount of computational power and energy~\cite{DBLP:conf/hicss/MenatiCHTX24}.

A cryptojacking attack allows an adversary to gain unauthorized access to a computer, server, or network of devices and installs malicious software on them~\cite{DBLP:journals/computer/KshetriV22}. This software aims to use the computing resources of the compromised systems for cryptocurrency mining. This activity takes place without the owners' knowledge or permission. The cryptocurrency industry has, unfortunately, been plagued by malicious activities, and cryptojacking remains a significant threat~\cite{DBLP:journals/corr/abs-2311-14783}. Due to the high costs of establishing and maintaining cryptocurrency infrastructures, attackers frequently exploit platforms owned by others to carry out their nefarious activities~\cite{DBLP:journals/ieeesp/CarlinBOS20}. By commandeering existing infrastructures, they can launch attacks against their desired targets without incurring any of the associated expenses.

The process of cryptocurrency mining is both intricate and demanding, calling for substantial investment. Unfortunately, potential profits often tempt attackers who actively search for systems that can generate the greatest returns~\cite{DBLP:journals/access/Hajiaghapour-Moghimi23}. In the past, private servers---which typically consume vast amounts of energy---were primarily used for mining. Nowadays, cloud-based servers offer more accessible, cost-effective options. However, some miners will hijack others' infrastructures to maximize their profits.
 
This paper aims to analyze and understand the infrastructure of certain websites previously linked to cryptojacking activities. We aim to determine whether these websites remain malicious through cryptojacking malware or other malicious software. We also seek to understand the geographical distribution of these sites and extract additional information to gain insights into their operations. Furthermore, we will discuss the security risks and threats associated with cryptojacking infrastructures and activities.

\section{Related Work}\label{sec:related} 
Numerous studies have explored different facets of cryptojacking~\cite{SaadSNKSNM20}. Jayasinghe~\etal~\cite{DBLP:conf/apit/JayasingheP20} explore cryptojacking within public cloud infrastructures, providing insights into how these environments are exploited for illicit cryptocurrency mining. Meanwhile, Saad~\etal~\cite{DBLP:journals/corr/abs-1809-02152} analyze end-to-end in-browser cryptojacking, examining how cryptojacking scripts operate within web browsers. Both studies significantly inform our understanding of cryptojacking across different platforms and contexts. Burgess \etal\ \cite{595ab4b3fc5c491dba0ba9a1820f4eb6} developed MANiC (Multi-step Assessment for Crypto-miners), a system designed to detect cryptocurrency mining scripts. MANiC extracts parameters that can be used to identify suspicious behaviors associated with mining activities, although it does not focus on web infrastructure. Carlin \etal\ \cite{5c91afde0b1242f1b0d632918f209e13} conducted a similar study focused on detecting cryptojacking websites by dynamic opcode analysis.
Xiao \etal \cite{DBLP:conf/mobicom/0002LR0H23} explored GPU cryptojacking and introduced MagTracer, a detection system for GPU-based cryptojacking. MagTracer boasts a detection accuracy of 98\%.

Naseem \etal \cite{DBLP:conf/ndss/NaseemABTU21} proposed MINOS, a lightweight detection system designed to identify cryptojacking activities in real time. Tekiner \etal\ \cite{DBLP:conf/ndss/TekinerAU22} presented an approach for detecting cryptojacking in IoT environments. Saad \etal\ \cite{DBLP:journals/corr/abs-2304-13253} conducted an in-depth analysis of cryptojacking samples, focusing on content, currency, and code-based categorization. Rajba \etal\ \cite{DBLP:conf/IEEEares/RajbaM22} presented an analysis highlighting the limitations of web cryptojacking detection methods. Lachtar \etal \cite{DBLP:journals/cal/LachtarEBM20} discussed a cross-stack approach to defending against cryptojacking. 

\BfPara{Research Gap} Our research paper distinguishes itself from others by focusing on websites engaged in cryptojacking. While other studies concentrate on detecting these websites, we thoroughly analyze the infrastructure, including their geographic distribution and other pertinent characteristics. Additionally, we perform a follow-up scan to identify any malicious content or malware and reclassify them as benign or malicious.

\section{Problem Statement and Research Questions}\label{sec:rqs} 
Generating a new cryptocurrency involves mining~\cite{DBLP:journals/corr/abs-1812-08806}. This complex process includes solving intricate mathematical problems that ensure the validation and security of transactions on the blockchain network~\cite{10328188}. The process of mining cryptocurrency differs from one type to another, and using consensus algorithms, such as Proof of Work (PoW) used by Bitcoin, and Proof of Stake (PoS), which determine how new blocks are added to the blockchain~\cite{DBLP:journals/corr/abs-2303-06008}.  

Mining cryptocurrencies demands substantial computational power, particularly in PoW systems. Miners must have access to robust hardware to solve intricate mathematical problems, which has led to the emergence of cryptojacking. We have analyzed the components and properties of cryptojacking infrastructures to gain insight into their operations. Our goal is to address the challenges posed by these infrastructures by answering three crucial research questions.

\begin{enumerate}
	\item {\bf RQ1: What are the underlying affinities between websites and malicious content, and how can these affinities be measured to assess the potential for cryptojacking activities?}. This enables us better to understand the relationship between the websites and malicious content.

 	\item {\bf RQ2: what categories of malicious content and malware are common in these cryptojacking infrastructures?} We analyzed various websites and categorized different types of malicious content and malware commonly associated with crypto-jacking activities. Our findings can provide valuable insights into the different categories of threats.

	\item {\bf RQ3: What are the hosting patterns and the geographical distribution of cryptojacking infrastructure?} The identified malicious websites are hosted in various countries globally. Analyzing this distribution will provide insights into the geographical spread of these sites.

\end{enumerate}

\section{Technical Approach}\label{sec:approach}
This study analyzed websites utilizing mining scripts for cryptocurrency mining to gain insights into the nature of cryptojacking infrastructures. It aimed to explore potential correlations between cryptojacking and other malicious activities, such as phishing campaigns~\cite{Mohaisen15}, malware distribution networks~\cite{MohaisenAM15,MohaisenA13,DuASAM23,AlrawiZDKLS19}, or botnets~\cite{WangCCM18}. The analysis began with the {\tt Whois} tool to gather comprehensive domain information, including IP addresses, name servers, and registrars.

We obtained a dataset of 887 websites associated with cryptocurrency mining from the MANiC dataset and subjected them to a comprehensive scan for malicious content using {\tt VirusTotal}. Out of the 887 websites, we successfully scanned approximately 880 of the websites. The results revealed that 371 websites were clean, while 518 contained malicious content. Our objective is to analyze the 518 websites detected to contain malicious content. This analysis aims to extract valuable information regarding the types of malicious content, the security engines that identified the threats, and the specific categories classified as malicious or suspicious. By decisively examining these factors, we assert our commitment to understanding the nature and extent of the threats posed by these websites.

\subsection{Dataset and Preprocessing}\label{sec:dataset_crypto}

\BfPara{Associated Cryptojacking websites} This paper used data from the "MANiC: Multi-step Assessment for Crypto-miners" dataset, which is also referenced in the paper titled "Detecting Cryptomining through Dynamic Analysis." The data was initially collected from the top 1 million websites on Alexa, as indexed by Censys, in July 2018. The dataset consists of multiple files, and our analysis focused on the malicious content file to examine its specifics.

After collecting the data in 2018, we conducted a re-scan to identify malicious content. Our re-scan revealed that 518 websites now contained malicious content. We utilized a domain query service to gather comprehensive information about all the websites in the dataset. This service allowed us to extract detailed domain information and other relevant data. Subsequently, we scanned the entire dataset to differentiate between infected and clean websites. This process aimed to enhance our understanding of the current threat landscape associated with cryptojacking and other malicious activities. 

\BfPara{Security Data Attributes} Afterward, we further analyzed the websites in the dataset with malicious infections by scanning them with VirusTotal. This analysis allowed us to obtain information about the security engines performing the scans and categorize the malicious detections into distinct groups. The output from the security engines is classified into three components: method, category, and result. VirusTotal uses the blacklist method to identify potentially harmful websites with malicious content~\cite{MohaisenA14x}. The security engine category can indicate suspicious or malicious activity, and the scan results can classify the content as malicious, suspicious, malware, phishing, or spam. We further analyzed the different types of malware and malicious content detected by each security engine. Various categories of malicious content were classified based on their threat levels and purposes. It is important to note that not all malicious content is related to cryptojacking activity. This classification helps us identify and concentrate on the malware and other malicious content that may be directly associated with cryptocurrency mining. Through this detailed analysis, we aimed to understand the types of threats these websites pose and differentiate between cryptojacking-related activities and other forms of malicious content.
 
\subsection{Analysis Dimensions}

This study investigates the infrastructure of cryptojacking by analyzing related websites. The goal is to identify malicious content and assess the current status of these websites. The research addresses specific questions outlined in section~\ref{sec:rqs} and covers various analytical dimensions.

\BfPara{\ding{182} Cryptojacking infrastructure websites distribution} This dimension explores the geographic and hosting distribution of websites engaged in cryptojacking. Analyzing the locations and hosting providers allows us to detect trends and hotspots in the infrastructure supporting cryptojacking activities.

\BfPara{\ding{183} Threat categories and classifications} Here, we analyze and categorize the threats from cryptojacking websites, focusing on identifying malicious behaviors like malware distribution, phishing, and unauthorized cryptocurrency mining.

\BfPara{\ding{184} Categorization of cryptojacking websites (current) into as malicious and benign} We analyze and categorize the current status of websites engaged in cryptojacking activities, distinguishing between those that are actively malicious, potentially harmful, or benign (no longer active or posing a threat).

\BfPara{\ding{185}  Malicious contents in the cryptojacking infrastructures} We investigate the specific types of malicious content present in cryptojacking websites' infrastructure, analyzing scripts, software, and techniques used for cryptojacking and other malicious activities. Our study offers a comprehensive analysis of cryptojacking infrastructure, providing insights into its distribution, threat landscape, current status, and the nature of malicious content it hosts.

\section{Results}\label{sec:results}

\subsection{Cryptojacking malware in the malicious dataset}
\begin{wrapfigure}{r}{0.5\textwidth}\vspace{-5mm}
    \centering
    \begin{tikzpicture}
        \begin{axis}[
            ybar,
            ymin=0, ymax=700,
            ytick distance=100,
            x=0.6cm,
            width=0.5\columnwidth, 
            height=0.3\columnwidth, 
            symbolic x coords={CoinHive, CryptoLoot, JSECoin, DeepMiner, Unknown, Webminepool, Papoto, CoinErra},
            xtick=data,
            xticklabel style={font=\small,rotate=45,anchor=east, align=right}, 
            yticklabel style={font=\small}, 
            grid=both,grid style={line width=.1pt, draw=gray!10},major grid style={line width=.1pt,draw=black!10},
            nodes near coords, 
            nodes near coords align={vertical}, 
        ]
        \addplot[color=black,pattern=crosshatch] coordinates {(CoinHive,678) (CryptoLoot,93) (JSECoin,92) (DeepMiner,13) (Unknown,19) (Webminepool,18) (Papoto,5) (CoinErra,1)};
        \end{axis}
    \end{tikzpicture}\vspace{-4mm}
\caption {This figure shows the prevalence of various cryptojacking malware types found within the malicious dataset, with CoinHive being the most significant contributor}
\label{fig:crypmal}\vspace{-6mm}
\end{wrapfigure}
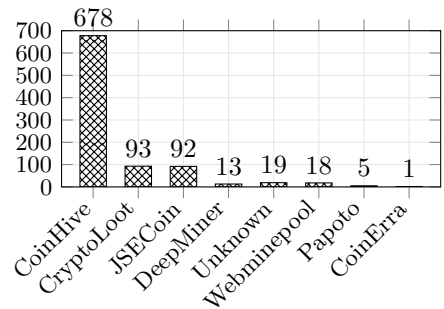

We analyzed the 919 websites included in the malicious file of the MANiC dataset. Our detailed examination revealed eight distinct types of cryptojacking malware within the dataset, as shown in \autoref{fig:crypmal}. One of the most significant cryptomining services identified in our analysis was CoinHive, which accounted for approximately 74\% of the malicious presence within the dataset. CoinHive was initially created as a legitimate service for cryptocurrency mining, enabling website owners to monetize their content by using visitors' CPU resources for mining. However, cybercriminals hijacked CoinHive for cryptojacking purposes, embedding its scripts into websites without users' consent to exploit their computing power for mining. Our findings indicate that even after its discontinuation (since 2019), the legacy of CoinHive continues to influence the prevalence and distribution of cryptojacking activities. This enduring impact is evident in the continued presence and proliferation of similar cryptomining malware. These insights underscore the need for vigilance and robust cybersecurity measures to mitigate the threats posed by such malicious activities.

\subsection{Cryptojacking infrastructure websites distribution} This study extensively analyzes numerous cryptojacking websites hosted across diverse geographic locations. Understanding the correlation between these locations is crucial for two main reasons. Firstly, examining the geographic distribution enhances our understanding of these websites' locations and the specific countries hosting them. This mapping helps identify trends and patterns that highlight regional vulnerabilities and the prevalence of cryptojacking infrastructure. Secondly, the analysis provides insights into how technological advancements and internet penetration contribute to cryptojacking activities. By correlating the presence of cryptojacking websites with regional technological landscapes, we can determine if higher technological development and internet usage correlate with increased cryptojacking incidents. This investigation also explores whether factors like broadband availability and digital literacy influence the spread of cryptojacking.

Addressing these questions will aid in identifying regions most impacted by cryptojacking. This data is crucial for devising strategies to combat cryptojacking, guiding policy decisions, and allocating resources to mitigate threats in vulnerable areas. Implementing effective discovery practices will enhance vigilance and security measures to safeguard networks against these pervasive threats. Of the 518 websites identified with malicious content, 116 were redacted for undisclosed reasons, limiting the analysis of country distribution to the remaining unredacted sites. Analyzing these sites allows for a focused exploration of geographical trends and patterns associated with malicious online activities.

\begin{figure}
    \centering\vspace{-4mm}
    \begin{tikzpicture}
        \begin{axis}[
            ybar,
            ymin=0, ymax=200,            ytick distance=100,
            x=0.6cm,
            width=0.5\columnwidth, 
            height=0.25\columnwidth, 
            symbolic x coords={USA, Iceland, India, China, Brazil, GB, Canada, St-Kitts,France,Seychelles,Russia,Bahama,Australia,Indonesia,Germany},
            xtick=data,
            xticklabel style={font=\small,rotate=45,anchor=east, align=right}, 
            yticklabel style={font=\small}, 
            grid=both,grid style={line width=.1pt, draw=gray!10},major grid style={line width=.1pt,draw=black!10},
            nodes near coords, 
            nodes near coords align={vertical}, 
        ]
        \addplot[color=black,pattern=crosshatch] coordinates {(USA,188) (Iceland,48) (India,18) (China,13) (Brazil,13) (GB,12) (Canada,7) (St-Kitts,7) (France,7) (Seychelles,6) (Russia,6) (Bahama,6) (Australia,5) (Indonesia,5) (Germany,4)};
        \end{axis}
    \end{tikzpicture}\vspace{-4mm}
\caption[Cryptojacking websites distribution.]{Cryptojacking websites distribution. A heavy-tailed distribution regarding the number of websites associated with cryptojacking activities.}
\label{tab:countries}\vspace{-4mm}
\end{figure}
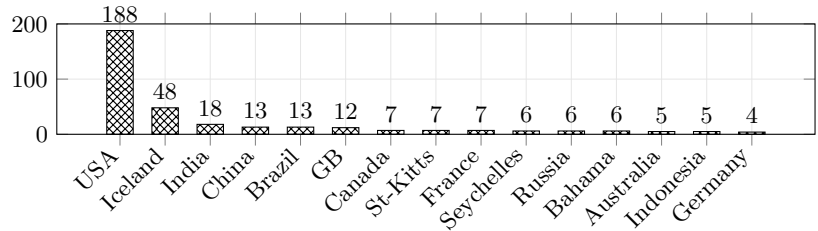

\BfPara{Observations} 
\autoref{tab:countries} displays the top 15 countries with the most websites infected with cryptojacking malware, as per our dataset. The websites are spread out in various countries across the Americas, Asia, South America, and Europe. We observed that the countries are advanced and have high internet penetration. The USA has more than 55\% of the websites, followed by Iceland, which has about 15\%. The USA is known to be well-advanced technologically and hosts most of the domains. The prevalence of cryptojacking websites in the United States may be attributed to the country's significant number of registered websites. Cryptocurrency has gained significant global recognition in recent years. However, it is notable that most cryptocurrency miners reside in regions with high internet penetration and reliable electricity access. Despite the widespread popularity of cryptocurrency investment and trading, the concentration of mining activities in such regions highlights the importance of favorable infrastructure for this practice. This observation suggests a correlation between the two variables and warrants further investigation.

\subsection{Threat categories and classifications}  
\begin{figure}\vspace{-5mm}
    \centering
    \begin{tikzpicture}
        \begin{axis}[
            ybar,
            ymin=0, ymax=200,
            ytick distance=100,
            x=0.6cm,
            width=0.5\columnwidth, 
            height=0.25\columnwidth, 
            symbolic x coords={alphaMountain.ai, Sophos, Fortinet, Webroot, Seclookup, ThreatSeeker, Scumware.org, PreCrime,CyRadar,Xcitium,Avira,CRDF,Quttera,Heimdal-Security,Safebrowsing},
            xtick=data,
            xticklabel style={font=\small,rotate=45,anchor=east, align=right}, 
            yticklabel style={font=\small}, 
            grid=both,grid style={line width=.1pt, draw=gray!10},major grid style={line width=.1pt,draw=black!10},
            nodes near coords, 
            nodes near coords align={vertical}, 
        ]
        \addplot[color=black,pattern=crosshatch] coordinates {(alphaMountain.ai,187) (Sophos,87) (Fortinet,54) (Webroot,39) (Seclookup,39) (ThreatSeeker,35) (Scumware.org,14) (PreCrime,8) (CyRadar,7) (Xcitium,6) (Avira,6) (CRDF,6) (Quttera,5) (Heimdal-Security,4) (Safebrowsing,3)};
        \end{axis}
    \end{tikzpicture}\vspace{-4mm}
\caption{Security engines categories with the number of occurrences.}
\label{tab:crypto}\vspace{-4mm}
\end{figure}
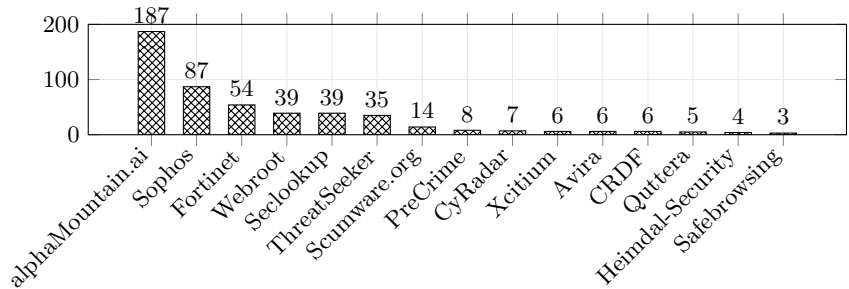

VirusTotal is the primary tool we use to scan malicious websites. After conducting scans, we detected various malicious activities, such as phishing, malware, and spyware identified by different security engines. We focused on these security engines and their reported malicious content to gain deeper insights into the nature of the threats. While the direct results from VirusTotal did not explicitly indicate crypto mining-related scripts, further analysis of the referrer files revealed the presence of such scripts. This was done by examining the code and behavior of the files, looking for patterns commonly associated with cryptomining scripts. The scan identified a total of 27 security engines, out of which 7 reported the presence of malicious content. \autoref{tab:sec_engine} summarizes the security engines and the corresponding malicious and suspicious contents they detected. Moreover, \autoref{tab:crypto} contains the list of the top 15 security engines and the number of malicious contents, including malicious, suspicious, and undetected contents.

\begin{table}[t]
\caption{Security vendors with the threat types.}
\label{tab:sec_engine}
\begin{tabular}{l|p{0.69\linewidth}}
\xl{2}
\textbf{Vendor} & \textbf{Threat type}\\
\xl{1}
Forcepoint ThreatSeeker & media file, compromised website, proxy avoidance, application and software download, hacking, potentially unwanted, suspicious content, p2p file sharing, and uncategorized. \\
\hline
Dr.Web & adult content, known infection source, gambling. \\
\hline
Webroot & malware, p2p, Bitdefender, porn, proxy avoidance and anonymizers, phishing and other frauds, spyware, and adware. \\
\hline
alphaMountains.ia & suspicious, malicious, unrated, anonymizers, JSEcoin, scam, illegal, unethical, coin\-hives, piracy, plagiarism. \\
\hline
Sophos & spyware and malware, pua and others, phishing and fraud, suspicious, spam URLs, proxies. \\
\hline
XcitiumverdicCloud & media sharing, spyware, malware. \\
\hline
Bitdefender & proxies, file sharing, webproxy \\
\xl{2}
\end{tabular}\vspace{-5mm}
\end{table}

\BfPara{Observations} 
The outcome of a security scan is determined by the security engine's ability to accurately identify and classify a malicious item. Sometimes, it may give false positives. It is important to note that the scan results are based on a specific period and could change over time due to website or security engine updates to correctly identify undetected or misclassified files.

\subsection{Malicious contents in the cryptojacking infrastructures} 
We organize the results from the VirusTotal scan into four distinct categories: harmless, malicious, suspicious, and undetected. Harmless files and contents are benign, posing no direct or indirect threat to users or systems. The scan results presented in Table \ref{tab:web-categ} indicate that most detected files fall into the harmless category, suggesting they are safe for use without any security concerns.

On the contrary, malicious file content can be hazardous and potentially harmful. These malicious contents can be injected into infrastructure for various nefarious purposes, including cryptocurrency mining. The potential harm of these contents underscores the importance of security measures and the need for caution. Cryptocurrency mining involves leveraging website resources for mining purposes without the explicit consent of website owners or visitors.

Additionally, malicious files can disrupt websites, temporarily unavailable or hijacking them until a ransom is paid~\cite{AlrawiIPKBHHS21}. The potential impact of these files on websites underscores the severity of the threat and the need for robust security measures. Such files may contain malware, spyware, or phishing components to compromise user data or system integrity.

\begin{table}[t]
\centering\vspace{-3mm}
\caption{The outcomes of security engines' classification.}
\label{tab:web-categ}
\begin{tabular}{p{1.5cm}p{1.5cm}p{1.5cm}p{1.5cm}p{1.5cm}}
\xl{2}
 Category & Harmless & Malicious & Suspicious & Undetected \\
\xl{1}
Count & 35,370 & 948 & 395 & 8,394 \\
\xl{2}
\end{tabular}\vspace{-4mm}
\end{table}

We analyzed 887 websites and found that 58\% are malicious and contain either suspicious or malicious files. The remaining 42\% are harmless or have undetected files. For further analysis, \autoref{tab:mal} summarizes the malicious websites. We categorize the files into three: benign, suspicious, and malicious. These websites contain benign or harmless files, but the presence of malicious and suspicious files qualifies them as malicious websites.

\begin{table}
\centering\vspace{-7mm}
\caption{Malicious contents and their distribution.}
\label{tab:mal}
\begin{tabular}{p{3.8cm}p{1.5cm}p{1.8cm}p{1.5cm}}
\xl{2}
Security Engines & Benign & Suspicious & Malicious  \\
\xl{1}
Forcepoint
  ThreatSeeker & 21              & 10                  & 4                   \\
\hline
Dr.
  Web                 & 0               & 2                   & 1                   \\
\hline
Webroot                   & 0               & 1                   & 5                   \\
\hline
alphaMountain.ia          & 7               & 4                   & 3                   \\
\hline
Sophos                    & 7               & 7                   & 5                   \\
\hline
XcitiumVerdictCloud       & 1               & 0                   & 2                   \\
\hline
BitDefender               & 12              & 5                   & 0    \\   
\xl{2}
\end{tabular}\vspace{-4mm}
\end{table}

\BfPara{Observations} \autoref{tab:mal} data clearly shows that the number of benign and suspicious files significantly exceeds that of malicious files. The table visually represents the distribution of files according to their threat levels. Notably, benign files, which are not considered harmful, comprise most of the files analyzed. These files pose no threat to the system and are safe for use. Following benign files, suspicious files constitute the next largest group. While these files are not confirmed to be harmful, they exhibit behaviors or characteristics that warrant further investigation to determine their true nature and potential threat level.
In contrast, the number of malicious files known to be harmful and pose a significant risk to the system is comparatively smaller. The presence of these files can cause various types of damage, including data breaches, system corruption, and unauthorized access. This distribution highlights the dataset's predominance of non-malicious files while scrutinizing suspicious files to ensure system security and mitigate potential threats.

\subsection{Correlation and statistical test}
To understand the correlation between malicious, suspicious, and undetected content on the website and conduct statistical analyses, we conducted two specific statistical tests: one between malicious and suspicious contents and another between malicious and undetected contents. The objective of these tests is to verify the existence of any relationship between the contents and their potential effects. Our primary aim is to identify the contributing factors to the presence of malicious activity on the websites. \autoref{tab:web-categ} and \autoref{tab:mal} show the count of the threat classifications and the website categories. The Pearson correlation coefficient data is based on the frequency of malicious and suspicious content on 517 malicious websites. This dataset enables us to measure and examine the linear relationship between the presence of malicious and suspicious content on these websites. The alpha ($\alpha$) for the null hypothesis is chosen as 0.05

\BfPara{Malicious and suspicious content in the websites} The Pearson correlation coefficient is a quantitative statistical measure utilized to assess the presence and strength of a linear correlation between malicious and suspicious content. It quantifies the degree to which the two variables are linearly related, providing valuable insights into the extent of their association.  From \autoref{tab:mal-susp}, The Pearson correlation coefficient indicated a small but significant negative relationship between malicious and suspicious content. The negative coefficient suggests that changes in these two variables occur in opposite directions. The difference between the malicious and suspicious content is substantial enough to be statistically significant. Due to the significance and the high $p$-value, we rejected the null hypothesis because there is a relationship between the two datasets.

\BfPara{Malicious and undetected content in the websites} We conducted a Pearson correlation analysis between malicious and undetected content using a dataset of 517 websites to explore their potential relationship. From \autoref{tab:mal-undet}, the analysis reveals a significantly positive correlation between these variables. This positive correlation suggests that changes in undetected content correspond proportionally with changes in malicious content in the same direction. Despite a relatively small difference, the correlation is statistically significant, supported by a sufficiently low $p$-value to reject the null hypothesis. This finding indicates that an increase in undetected content is associated with a proportional increase in malicious activities on these websites.

    \begin{table}[t]
    \begin{minipage}{.32\textwidth}
    \centering
        \caption{Correlation between malicious and suspicious content.}
    \label{tab:mal-susp}
    \begin{tabular}{ll}
    \xl{2}
    \textbf{Parameter} & \textbf{Value} \\ 
    \xl{1}
    Coefficient ($r$) & -0.2143 \\ 
    $r^2$ & 0.04594 \\ 
    $p$-value & 8.696e-7 \\ 
    Covariance & -0.3048 \\ 
    Sample   size ($n$) & 517 \\ 
    Statistic & -4.9799 \\ 
    \xl{2}
    \end{tabular}\vspace{-7mm}
    \end{minipage}
    \begin{minipage}{.32\textwidth}
    \centering
    \caption{Correlation between malicious and undetected content.}
    \label{tab:mal-undet}
    \begin{tabular}{ll}
    \xl{2}
    \textbf{Parameter} & \textbf{Value} \\ \xl{1}
    Coefficient ($r$) & 0.2335 \\ 
    $r^2$ & 0.05452 \\ 
    $p$-value & 7.834e-8 \\ 
    Covariance & 1.9366 \\
    Sample   size ($n$) & 517 \\ 
    Statistic & 5.4497 \\ \xl{2}
    \end{tabular}\vspace{-7mm}
    \end{minipage}~
    \begin{minipage}{.32\textwidth}\centering
        \caption{Correlation between malicious and suspicious content.}
    \label{tab:mal-sus-sec}
    \begin{tabular}{ll}
    \xl{2}
    \textbf{Parameter} & \textbf{Value} \\ \xl{1}
    Coefficient ($r$) & 0.2694 \\ 
    $r^2$ & 0.07257 \\ 
    $p$-value & 0.5591 \\ 
    Covariance & 1.8571 \\ 
    Sample   size ($n$) & 7 \\ 
    Statistic & 0.6255 \\ \xl{2}
    \end{tabular}\vspace{-7mm}
    \end{minipage}
    \end{table}

\BfPara{Suspicious and malicious content from the security engines} We investigate the correlation between malicious and suspicious content detected by security engines. The Pearson correlation coefficient from \autoref{tab:mal-sus-sec} indicates a non-significant positive relationship between these variables. This slight positive correlation suggests that changes in one variable have minimal impact on the other, indicating little mutual influence. Due to the small and statistically insignificant difference, we cannot reject the null hypothesis that there is no correlation between suspicious and malicious content as identified by the security engines. Therefore, the small $p$-value does not provide enough evidence to support a meaningful relationship between these variables.

\subsection{Summary of Results and Findings}\label{sec:results_crypto}

While some content is flagged as suspicious or undetected, it may contain elements contributing to the websites' malicious nature. Some of the suspicious and undetected files likely harbor hidden malicious content, which suggests that the security scanning software may not have correctly classified them. Collectively, this increases the number of malicious instances on these websites. Therefore, examining and reclassifying suspicious and undetected files is imperative to understand better and mitigate their contribution to malicious activity. Additional findings are illustrated below:

\BfPara{Correlation between website contents} Our analysis reveals a correlation between websites hosting malicious and suspicious content, such as phishing links, malware, or suspicious scripts, and those with malicious and undetected content. This correlation suggests that websites exhibiting such content may harbor malicious elements. Understanding these correlations is crucial for effectively identifying and mitigating cyber threats.
    
\BfPara{Geographical concentration of cryptojacking websites} Most cryptojacking websites are concentrated in a few countries with high internet penetration. We can attribute this concentration to various factors, including lax cybersecurity regulations, widespread use of outdated software, and a need for more awareness among website owners and users. Addressing this concentration requires coordinated efforts to raise awareness about the risks of cryptojacking.

\BfPara{Redacted domain information} Many domains exhibit redacted or ``unavailabl'' information, challenging ascertaining ownership and origins. This lack of transparency hinders tracking and addressing malicious activities associated with these domains. Improving transparency and accountability in domain registration processes is necessary to enhance cybersecurity measures.
    
\BfPara{Observation of malicious contents} We identified several instances of malicious content associated with cryptojacking. These contents, including scripts and malware, exploit vulnerabilities in websites for illicit mining using visitors' computational resources. Detecting and removing such malicious content is critical for safeguarding users' security and privacy~\cite{DuanBJAXISL19,PerdisciPAA20}.
    
\BfPara{Cleanup of previously malicious websites} Many websites previously flagged as malicious have undergone cleanup, removing the malicious content. This cleanup indicates proactive measures website owners or security professionals take to mitigate the impact of cryptojacking activities. However, continuous monitoring is necessary to prevent reoccurrences of such incidents.
   
\BfPara{Return of previously malicious websites to benign status} Some previously identified as malicious websites have returned to a benign state, indicating a potential temporary hijacking for cryptomining purposes. This phenomenon underscores the dynamic nature of cyber threats, where threats can evolve and change over time, and the importance of timely detection and response mechanisms to thwart cryptojacking attempts effectively.
    
\BfPara{Absence of correlation between malicious and suspicious contents in security engines} Our analysis indicates no correlation between malicious and suspicious contents in security engines. This finding suggests that while security engines may flag specific contents as suspicious, they may not necessarily classify them as malicious. Understanding this distinction is crucial for refining threat detection algorithms and improving the accuracy of security assessments.

\section{Concluding Remarks} \label{sec:conclusion}
We investigated websites compromised for cryptocurrency mining. Using the whois tool, we identified the geographic distribution of these websites, primarily located in regions with significant internet usage. According to our analysis using \url{virustotal.com}, out of the 887 websites previously classified as malicious, 370 no longer contain malicious content or crypto-jacking scripts. This improvement may stem from enhanced website security measures or reduced attractiveness to attackers. However, some sites still harbor crypto-jacking-related malware, clandestinely exploiting users' computing resources for unauthorized cryptocurrency mining. Our findings suggest these websites remain susceptible to re-infection. We acknowledge the limitations of our study and emphasize the need for future research to provide deeper insights into cryptojacking dynamics and lifecycle.


\end{document}